\documentclass[a4paper,twoside,11pt]{article}
\usepackage{amssymb}
\usepackage{amsmath}
\usepackage{amsfonts}

\begin{document}

\title{\Large\bf{Generalized Tomonaga--Schwinger equation\\
from the Hadamard formula}} \author{{\normalsize{Luisa Doplicher}}
\\[1mm] \em \small{\!\!\! Dipartimento di Fisica dell'Universit\`a
``La Sapienza", INFN Sez.\,Roma1, I-00185 Roma, EU}}
\date{{} }

\maketitle

\begin{abstract}

A generalized Tomonaga--Schwinger equation, holding on the entire
boundary of a {\em finite} spacetime region, has recently been
considered as a tool for studying particle scattering amplitudes
in background-independent quantum field theory.  The equation has
been derived using lattice techniques under assumptions on the
existence of the continuum limit.  Here I show that in the context
of continuous euclidean field theory the equation can be directly
derived from the functional integral formalism, using a technique
based on Hadamard's formula for the variation of the propagator.

\end{abstract}

\section{Introduction}

Quantum dynamics can be formulated in a variety of ways.  One of
these, championed by Feynman \cite{feynhibbs}, is to give the
propagator of the Schr\"odinger evolution $W(q_{1},q_{2},\,
t)$.  Here $q_{1}$ and $q_{2}$ are initial and final positions
and $t$ is the propagation time.  The propagator satisfies the
Schr\"odinger equation in the initial as well as in the final
variables.  Formally, it can be expressed as the path integral of the
exponential of the action, over all paths that start at $q_{1}$ at
time zero and end at $q_{2}$ at time $t$
\begin{equation}
W(q_{1},q_{2},\, t)= \int_{\stackrel{\scriptscriptstyle
x(0)=q_{1}}{\scriptscriptstyle x(t)=q_{2}}} Dx\
e^{-\frac{i}{\hbar} S[x]}\ .
\end{equation}%
Quantum field theory too can be formulated in a variety of ways.
One of these is the Schr\"{o}dinger functional representation
(for a review, see \cite{jackiw}), where states are expressed as
wave functionals of the field $\varphi$ defined on a spacelike 3d
plane. Evolution is given by a functional-differential
Schr\"{o}dinger equation.  The propagator of this equation is a
functional $W[\varphi_{1},\varphi_{2},\, t]$, which propagates
the initial field $\varphi_{1}$ at time zero to the final field
$\varphi_{2}$ at time $t$.  I give the explicit form of
$W[\varphi_{1},\varphi_{2},\, t]$ for a free field theory in an
appendix below.  The functional $W[\varphi_{1},\varphi_{2},\, t]$
satisfies the functional Schr\"odinger equation on both the
initial and the final spacelike 3d planes.  Formally, it can be
expressed as the path integral of the exponential of the action,
over all field configurations that take the initial value
$\varphi_{1}$ on the initial plane and take the final value
$\varphi_{2}$ on the final plane.

Robert Oeckl \cite{oeckl} has recently argued that it should be
possible to formulate quantum field theory in terms of a
functional $W[\varphi, \Sigma]$ of the field $\varphi$ defined on
the compact 3d boundary $\Sigma$ of a {\em finite} spacetime
region $\cal R$.  In this case, $\Sigma$ is not going to be
spacelike.  The same point of view has been extensively developed
by Carlo Rovelli in the book \cite{book}, to which I refer for a
detailed discussion.  Formally, $W[\varphi, \Sigma]$ can be
expressed as a path integral over all field configurations on
$\cal R$, that take the boundary value $\varphi$ on the 3d
boundary $\Sigma$
\begin{equation}
W[\varphi, \Sigma] =\int_{\left.  \Phi \right\vert_{\Sigma }=
\varphi}D\Phi\  \   e^{-\frac{i}{\hbar} S[\Phi]}
\ .  \label{K}
\end{equation}
The field $\varphi$ can be viewed as expressing initial, final as
well as boundary values of $\Phi$, and $W[\varphi, \Sigma]$
expresses the amplitude for having a certain set of initial and
final fields, as well as boundary fields, measured by devices
that are located in spacetime in the manner described by (the
geometry of) the surface $\Sigma$. This picture is near to what
actually happens in a laboratory experiment, where the initial
and final state of a scattering event are confined into
finite-size spacetime regions.  The relation between particle
states that can be defined in such a finite context and the usual
particle states of quantum field theory, defined on an infinite
spacelike region, is discussed in \cite{daniele}.

The interest of this suggestion is in its relevance for quantum
gravity.  Here, diffeomorphism invariance (that is, background
independence) implies that $W[\varphi, \Sigma]$ is independent
from $\Sigma$.  The key observation, pointed out in \cite{cdort},
is that the boundary value of the gravitational field on the
boundary surface $\Sigma$ determines the geometry of $\Sigma$,
and therefore codes the relative distance of the measuring
devices and the proper time elapsed from the bottom to the top of
the spacetime region considered. Therefore in gravity the {\em
background independent} propagator $W[\varphi]$ contains the same
physical information about the geometry of the boundary surface
as the propagator $W[\varphi, \Sigma]$ of a {\em background
dependent} quantum field theory.  In a scattering experiment we
measure incoming and outgoing particles, that is, matter-fields
variables, as well as distances between instruments and elapsed
time, that is, gravitational-field variables.  The two kind of
variables are both represented by the boundary fields $\varphi$.
In other words, the boundary formulation realizes very elegantly
in the quantum context the complete identification between
spacetime geometry and dynamical fields, which is the great
Einstein's discovery.  A simple model where this picture is
concretely realized is presented in \cite{cubo}.  Thus, it was
suggested in \cite{cdort} that particle scattering amplitudes can
be effectively extracted from $W[\varphi]$ in a background
independent theory; details will be given elsewhere
\cite{florian}.

Given the possible relevance of the propagator $W[\varphi,
\Sigma]$ for quantum gravity, it is important to investigate it
in detail, and to check whether its expected properties truly
hold.  In particular, it was argued in \cite{cdort} that
$W[\varphi, \Sigma]$ should satisfy a local functional equation
governing the variation of $W[\varphi, \Sigma]$ under arbitrary
local deformations of $\Sigma$, namely an equation of the form
\begin{equation}
\frac{\delta W[\varphi, \Sigma]}{\delta \Sigma(s)} =
H\!\left(\varphi(s), \nabla\phi(s), \frac{\delta}{\delta
\varphi(s)}\right)\ W[\varphi, \Sigma].
\label{equazione}
\end{equation}
Here $s$ is a coordinate on $\Sigma$, and the precise meaning of
the other symbols will be defined below.  When $\Sigma$ is formed
by two parallel planes, (\ref{K}) becomes the field propagator,
and its variation under a parallel displacement of one of the
planes is governed by the functional Schr\"odinger equation.  The
possibility of extending this equation to variations of arbitrary
{\em spacelike} 3d surfaces was explored by Tomonaga and Schwinger
already in the late forties \cite{tomschw}.  These authors
derived a well-known local generalization of the functional
Schr\"odinger equation, called the Tomonaga--Schwinger equation.
The functional equation
(\ref{equazione}) generalizes the Tomonaga--Schwinger equation,
since it holds for arbitrary boundary surfaces.
Indeed, if the general philosophy of the boundary approach is
correct, the distinction between the initial and final fields on
spacelike portions of $\Sigma$ on the one hand, and the boundary
values of the field on the timelike portions of $\Sigma$, should
become of secondary relevance.

Equation (\ref{equazione}) has been derived in \cite{cr} on the
basis of a lattice regularization of the functional integral
(\ref{K}) defining $W[\varphi, \Sigma]$, and under certain
hypotheses on the existence of the continuum limit.  Here,
working in the context of a free euclidean theory, I show that
this equation can be derived from the functional integral
definition of $W[\varphi, \Sigma]$ directly in the continuum,
using a formula by Hadamard which expresses the variation of a
Green function under variation of the boundary \cite{volterra}.
Although incomplete, I think that this is a relevant step towards
establishing the general viability of an equation of the form
(\ref{equazione}).

Furthermore, as a preliminary step in the derivation of
(\ref{equazione}), I derive its classical limit, which is a
generalized Hamilton--Jacobi functional equation of the form
\begin{equation}
\frac{\delta S[\varphi, \Sigma]}{\delta \Sigma(s)} =
H\!\left(\varphi(s), \nabla\phi(s), \frac{\delta S[\varphi,
\Sigma]}{\delta \varphi(s)}\right) \label{equazioneHJ}
\end{equation}
satisfied by the classical Hamilton function $S[\varphi, \Sigma]$.
The Hamilton function $S[\varphi, \Sigma]$ is the value of the action
computed on the solution of the equations of motion that takes the
value $\varphi$ on $\Sigma$ (see \cite{book}.)  The generalized
Hamilton--Jacobi equation (\ref{equazioneHJ}) is extensively discussed
in \cite{book}.

The paper is organized as follows.  In Section 2, I define the
theory and equation (\ref{equazione}); in Section 3 I introduce
some Green function techniques; in Section 4 I prove equation
(\ref{equazioneHJ}); in section 5 I prove equation
(\ref{equazione}). In the Appendices I sketch a proof of the
Hadamard formula, and I give the explicit expression of the
functional Schr\"odinger propagator
$W[\varphi_{1},\varphi_{2},\, t]$ for the euclidean and for the
lorentzian theory.

\section{Definitions}

\subsection{Surface and surface derivative}

Consider a finite region $\cal R$ in the euclidean 4d space
$R^4$.  I use cartesian coordinates $x, y, z \ldots$ on $R^4$,
where $x=(x^\mu),\ \mu=1,2,3,4$.  Let $\Sigma =\partial{\cal
R}_{\Sigma}$ be a compact 3d surface that bounds a finite region
${\cal R}_{\Sigma}$. I denote $s,t,u\ldots$ coordinates on
$\Sigma$, where $s=(s^a),\ a=1,2,3$.  Then $\Sigma$ is given by
the embedding
\begin{equation}
\Sigma: s^a \longmapsto x^\mu(s).
\label{Sigma}
\end{equation}
The euclidean 4d metric, which raises and lowers the $\mu$ indices,
induces the 3d metric
\begin{equation}
q_{ab}(s)= \frac{\partial x^\mu(s)}{\partial s^a}
\frac{\partial x_{\mu}(s)}{\partial s^b}
\end{equation}
on $\Sigma$, which can be used to raise and lower the three
dimensional $a$ indices.  The surface gradient $\nabla$ is defined as
\begin{equation}
\nabla^a=q^{ab}\frac{\partial}{\partial s^b}.
\end{equation}
The normal one-form to the surface is
\begin{equation}
\tilde n_{\mu}(s)=\epsilon_{\mu\nu\rho\sigma}
\frac{\partial x^\nu(s)}{\partial s^1}
\frac{\partial x^\rho(s)}{\partial s^2}
\frac{\partial x^\sigma(s)}{\partial s^3}.
\label{normal}
\end{equation}
I orient the coordinate system $s$ so that $n_{\mu}$ is outward
directed.  Its norm is easily seen to be equal to the determinant
of the induced metric
\begin{equation}
\tilde n_{\mu} \tilde n^\mu = \det q.
\end{equation}
In the following, I will use the normalized normal
\begin{equation}
n_{\mu} \equiv (\det q)^{-\frac12}\  \tilde n_\mu
\end{equation}
and the induced volume element on $\Sigma$
\begin{equation}
d\Sigma(s) \equiv  (\det q(s))^{\frac12}  \ d^3s.
\end{equation}
(I shall simply write $d\Sigma$ when the integration variable is
clear from the context.)  Notice that the combination $d\Sigma\
n_{\mu}$ does not depend on the metric, since
\begin{equation}
d\Sigma(s)\ n_{\mu}(s) = d^3s\ \tilde n_{\mu}(s).
\end{equation}

Given a functional $F[\Sigma]$ that depends on the surface, I
define the functional derivative with respect to the surface as
the normal projection of the functional derivative with respect
to the embedding (\ref{Sigma}) that defines the surface
\begin{equation}
\frac{\delta }{\delta \Sigma (s) }\equiv
n^\mu(s)
\frac{\delta }{\delta x^\mu(s)}.
\label{derivata}
\end{equation}%
Here the functional derivative on the right hand side is defined in
terms of the volume element $d\Sigma$.  That is \cite{volterra},
${\delta F\left[ \Sigma \right] }/{\delta \Sigma (s) }$ is
defined as the distribution that satisfies
\begin{equation}
\int d\Sigma(s) \ N(s)\ \frac{\delta F\left[ \Sigma \right]
}{\delta \Sigma (s) }=
\int d\Sigma \ N(s)\ n^\mu(s)\
\frac{\delta  F\left[ \Sigma \right] }{\delta x^\mu(s)}
= \lim_{\varepsilon \rightarrow 0}
\frac{F\left[ \Sigma _{\varepsilon N }\right] -F\left[
\Sigma \right] }{\varepsilon }\  ,
\label{standardderivfunz1}
\end{equation}%
where $\Sigma_{\varepsilon N}$ is the surface defined by the embedding
\begin{equation}
\Sigma_{\varepsilon N}: s^a \longmapsto x^\mu(s)+\varepsilon N(s)
n^\mu(s).
\label{Sigmadef}
\end{equation}
Geometrically, this derivative gives the variation of the functional
under an infinitesimal displacement of the surface in the normal
direction.

The surface derivative takes a simple form for certain simple
functionals. Consider a scalar field $f(x)$ on $R^4$.
Let $F_{f}[\Sigma]$ be the integral of $f$ in the region ${\cal
R}_{\Sigma}$ bounded by $\Sigma$; that is
\begin{equation}
F_{f}[\Sigma] = \int_{\cal R} d^4x f(x) \  .
\end{equation}%
Then easily
\begin{equation}
\frac{\delta F_{f}[\Sigma]}{\delta \Sigma(s)} = f(x(s)).
\label{bulk}
\end{equation}%
That is, the variation of the bulk integral under a normal variation
of the surface is the integrand in the variation point.  Similarly,
consider a vector field $v^{\mu}(x)$ on $R^4$.  Let $F_{v}[\Sigma]$ be
the flux of $v^{\mu}(x)$ across $\Sigma$; that is
\begin{equation}
F_{v}[\Sigma] = \int_{\Sigma } d\Sigma(s)\ n_{\mu}(s)v^{\mu}(s) \  .
\end{equation}%
Then Stokes theorem gives easily
\begin{equation}
\frac{\delta F_{v}[\Sigma]}{\delta \Sigma(s)} =
\partial_{\mu}v^{\mu}(x(s)).
\label{flux}
\end{equation}%
That is, the variation of the flux by a normal variation of the
surface is the divergence of the vector field.

\subsection{Field theory and Tomonaga-Schwinger equation }

I consider a free euclidean scalar field $\Phi(x)$, defined in
$\cal R$, with assigned boundary conditions $\varphi(s)$ on
$\Sigma$. That is
\begin{equation}
\Phi(x(s)) = \varphi(s).
\label{boundary}
\end{equation}
The dynamics of the field is governed by the lagrangian density
\begin{equation}
\mathcal{L}=\frac{1}{2}\partial_{\mu}\Phi\partial^{\mu}\Phi
+\frac{1}{2}%
m^{2}\Phi ^{2}\  ,  \label{lagr}
\end{equation}%
and the equation of motion is
\begin{equation}
\left( -\square _{x}+m^{2}\right) \Phi \left( x\right) =0,\  x\in V
\label{eqmoto}
\end{equation}%
with the boundary conditions (\ref{boundary}) on $\Sigma $.  Here
$\square_{x}= \partial/\partial x_{\mu}\ \partial/\partial
x^{\mu}$. By choosing one of the coordinates $x^\mu$, say $x^4$ as
a ``time'' coordinate, it is possible to derive the hamiltonian
density
\begin{equation}
H(\Phi, \nabla \Phi,\Pi)=\frac{1}{2}\Pi^{2}-\frac{1}{2}%
\left( \nabla \Phi \right) ^{2}-\frac{1}{2}m^{2}\Phi ^{2}\  ;
\label{hamiltonian}
\end{equation}%
where $\Pi$ is the canonical momentum associated to the field $\Phi$
and $ \nabla$ is the gradient on the $x^4=constant$ surface.

For this theory, the integral in (\ref{K}) is a well defined
gaussian functional integral.  Therefore $W[\varphi, \Sigma]$ is
well defined. I show in this paper that $W\left[\varphi,
\Sigma\right]$ satisfies the generalized Tomonaga-Schwinger
equation
\begin{equation}
\begin{split}
\frac{\delta W[\varphi, \Sigma]}{\delta \Sigma (s) }&=
H\left(\varphi(s), \nabla \varphi(s), -\frac{\delta }{\delta
\varphi (s) }\right)\ W[\varphi, \Sigma]
\\
&=\left( \frac{1}{2}\left( -\frac{\delta }{\delta
\varphi(s)}\right)
^{2}-%
\frac{1}{2}\left( \nabla \varphi(s) \right)^{2}-\frac{1}{2}%
m^{2}\varphi ^{2}(s) \right) W[\varphi, \Sigma]\  .
\end{split}
\label{schr}
\end{equation}%

\section{Green functions}

The tool I use to prove (\ref{schr}) is a Green-function
technique. Given $\Sigma$, let the Green function $G_{\Sigma
}(x,y) $ be the
solution of the inhomogeneous equation%
\begin{equation}
\left( -\square _{x}+m^{2}\right) G_{\Sigma }\left( x,y\right) =\delta
^{\left( 4\right) }( x-y)  \label{defG}
\end{equation}%
in the region $\cal R$, satisfying the boundary condition%
\begin{equation}
G_{\Sigma }\left( x(s),y\right)=0\ ,
\label{GxeS=0}
\end{equation}%
namely that vanishes when $x$ is on $\Sigma$. I introduce the
following useful notation:
\begin{equation}
\partial_{n}^s G_{\Sigma }(s,y)\equiv
\left. n^\mu(s)\ \frac{\partial G_{\Sigma}(x,y)}{\partial x^\mu}
\right\vert_{x=x(s)}
\label{notation}
\end{equation}%
and similarly for $\partial_{n}^s G_{\Sigma }(y,s), \ \partial_{n}^s
\partial_{n}^t G_{\Sigma }(s,t), \ $ and so on.  The solution of (\ref
{eqmoto}), with the boundary condition (\ref{boundary}), namely with
the boundary value $\varphi$ on $\Sigma$, is then given by
\begin{equation}
\Phi_{\varphi,\Sigma}\left(y\right) =-\int_{\Sigma }d\Sigma(s) \
\varphi(s) \
\partial_{n}^{s}G_{\Sigma}(s,y)
\ .  \label{classica}
\end{equation}%
To see that this is indeed the solution, one can write the equality
\begin{equation}
\Phi \left( y\right) =\int d^{4}x\left( \Phi \left( x\right) \left(
-\square
_{x}+m^{2}\right) G_{\Sigma }\left( x,y\right) -G_{\Sigma }\left(
x,y\right)
\left( -\square _{x}+m^{2}\right) \Phi \left( x\right) \right) \  ,
\end{equation}%
which is satisfied because of (\ref{defG}) and (\ref{eqmoto}). The
right hand side can easily be written as a surface integral, which
gives (\ref{classica}) using the boundary conditions (\ref{boundary})
and (\ref{GxeS=0}).

The propagator $W[\varphi, \Sigma]$ can be written explicitly in
terms of the Green function.  In fact, it is possible to solve
the gaussian integral (\ref{K}) using the fact that the classical
solution (\ref{classica}) is an extremal value of the exponent in
(\ref{K}). The integral is then a gaussian integral over the
fluctuations.  The action on the classical solution with given
boundary conditions is called the Hamilton function
\cite{hamilton}:
\begin{equation}
S[\varphi, \Sigma] \equiv S[\Phi_{\varphi,\Sigma}] \ \ =\ \
\frac{1}{2}%
\int_{V_{\Sigma }}d^{4}x\left[ \left( \partial _{\mu
}{\Phi_{\varphi,\Sigma}}\right)
^{2}+m^{2}{\Phi_{\varphi,\Sigma}}^{2}\right] \quad . \label{S1}
\end{equation}%
Inserting (\ref{classica}), it is possible to write this as
\begin{equation}
S[\varphi, \Sigma] =-\frac{1}{2}\int_{\Sigma} d\Sigma(s)
d\Sigma(t) \ \varphi(s)\ \varphi(t) \
\partial_{n}^{s}\partial_{n}^{t}G_{\Sigma }(s,t) \  . \label{S}
\end{equation}%
It is then easy to perform the gaussian integration in (\ref{K}),
which gives
\begin{equation}
\begin{split}
W[\varphi, \Sigma] &=\frac{\exp \left( -S\left[
\Phi_{\varphi,\Sigma} \right] \right) }{\sqrt{\det \left(
-\square +m^{2}\right) }}
\\&= \sqrt{\det G_{\Sigma }\left( x,y\right) }\ \exp\left(
\frac{1}{2}\int_{\Sigma} d\Sigma(s) d\Sigma(t) \ \varphi(s)\
\varphi(t) \ \partial_{n}^{s}\partial_{n}^{t}G_{\Sigma
}(s,t)\right)\ . \
\end{split}
\label{KK}
\end{equation}

In the following, it is necessary to have an expression for the variations of Green function $%
G_{\Sigma }$ with respect to a displacement of the surface
$\Sigma $. This can be done using Hadamard's formula \cite{volterra}%
\begin{equation}
\frac{\delta G_{\Sigma }\left( x,y\right) }{\delta \Sigma (s) }%
=\partial _{n}^{s}G_{\Sigma }(s,x)\ \partial _{n}^{s}G_{\Sigma
}(s,y) \  .  \label{hadamard}
\end{equation}%
I sketch a proof of this formula in the Appendix A.

\section{Hamilton--Jacobi}

In this section I prove that the Hamilton function (\ref{S})
satisfies the generalized Hamilton--Jacobi equation
(\ref{equazioneHJ}).  From (\ref{hamiltonian}), this reads
\begin{equation}
\frac{\delta S[\varphi, \Sigma] }{\delta \Sigma (s) }+%
\frac{1}{2}\left( \frac{\delta S[\varphi, \Sigma] }
{\varphi(s) }\right) ^{2}-
\frac{1}{2}m^{2}\varphi^{2}(s)
-\frac{1}{2}\left( \nabla \varphi(s) \right) ^{2}=0\  .
\label{h-j}
\end{equation}%
To calculate the first term of this expression, I use the form
(\ref{S1}) of the Hamilton function
\begin{equation}
\frac{\delta S[\varphi, \Sigma] }{\delta \Sigma (s) }
=
\frac{\delta }{\delta \Sigma (s) }\ \
\frac{1}{2}%
\int_{{\cal R}_{\Sigma }}d^{4}x\left[ \left( \partial _{\mu
}{\Phi_{\varphi,\Sigma}}\right)
^{2}+m^{2}{\Phi_{\varphi,\Sigma}}^{2}\right]
\  .
\end{equation}%
I use (\ref{bulk}) for the variation of the integration region,
integrate by parts, use the equations of motion, to obtain
\begin{equation}
\frac{\delta S[\varphi, \Sigma] }{\delta \Sigma (s) }
=\frac{1}{2}%
\left( \left( \partial _{\mu }{\Phi_{\varphi,\Sigma}(x(s))}\right)
^{2}+m^{2}
{\Phi_{\varphi,\Sigma}(x(s))}%
^{2}\right)
+ \int_{{\cal R}_{\Sigma }}d^{4}x\ \partial _{\mu }\left( \partial
_{\mu }{\Phi_{\varphi,\Sigma}}%
\frac{\delta{\Phi_{\varphi,\Sigma}}}{\delta \Sigma (s) }\right)
\quad.
\end{equation}%
The second term can be written as a surface integral. Using
(\ref{classica}), this becomes
\begin{equation}
\begin{split}
\frac{\delta S[\varphi, \Sigma] }{\delta \Sigma(s) } =&
\frac{1}{2} \left( \left( \partial _{\mu }
{\Phi_{\varphi,\Sigma}(x(s))}\right)^{2}+
m^{2}{\Phi_{\varphi,\Sigma}(x(s))}^{2}\right)
 \\ &
-\int_{\Sigma }d\Sigma(t) d\Sigma(t) \ \varphi(t)\
\partial_{n}^{t}\partial_{n}^{u}G_{\Sigma }(t,u)
\frac{ \delta {\Phi_{\varphi, \Sigma}}(x(u)) }{\delta
\Sigma (s) }.
\end{split}
\label{meta}
\end{equation}%

To compute $\delta {\Phi_{\varphi,\Sigma}}\left( x\right) /\delta
\Sigma (s) $ observe that by defining the field $\varphi(x)$ in
the neighborhood of the surface by $\varphi(x(s))=\varphi(s)$ and
$\partial_{n}\varphi(x(s))=0$, the right hand side of
(\ref{classica}) is the flux of the vector field
\begin{equation}
v^\mu(x) =  -  \varphi(x) \ \frac{\partial}{\partial x^\mu}
G_{\Sigma }(x,y)
\  ,  \label{classicap}
\end{equation}%
Therefore it is possible to use (\ref{flux}) to obtain
\begin{equation}
\frac{\delta{\Phi_{\varphi,\Sigma}}\left( x\right) } {\delta \Sigma
(s) }= -\frac{\partial}{\partial y_\mu} \left( \varphi(y(s))
\frac{\partial}{\partial y^\mu}
G_{\Sigma }\left(y(s),x\right) \right) %
-\int_{\Sigma }d\Sigma(t)\ \varphi(t)\ n^{\mu}(t)\
\frac{\partial}{\partial y_\mu}
\frac{\delta G_{\Sigma }\left(y(t),x\right) }{\delta \Sigma(s)}
\end{equation}
In the first term, I separate the normal and tangential component
of the sum over $\mu$ and recall that $\varphi(x)$ is constant in
the normal direction, and in the second term I use the Hadamard
formula (\ref{hadamard}).  This gives
\begin{equation}
\begin{split}
\frac{\delta{\Phi_{\varphi,\Sigma}}\left( x\right) } {\delta
\Sigma (s) }=& -\nabla_{a}\varphi(s) \nabla_{a}G_{\Sigma
}\left(y(s),x\right) -\varphi(s) \square_{y}G_{\Sigma
}\left(y(s),x\right)  \\ &
 -\int_{\Sigma }d\Sigma(t)\  \varphi(t)\
n^\mu(t)\  \partial_{n}^{s}G_{\Sigma}(s,x)\
\partial_{\mu }^{t}\partial_{n}^{s}G_{\Sigma }(s,t) \ .
\end{split}
\end{equation}%
But the tangential derivative of the Green function on the
boundary surface vanishes, since the Green function itself
vanishes. So does its Dalambertian, since $\left( -\square
_{z}+m^{2}\right) G_{\Sigma }\left( z\in \Sigma ,y\right) =0$).
Therefore the first two terms vanish, and using  (\ref{classica})
again, the result is
\begin{equation}
\frac{\delta{\Phi_{\varphi,\Sigma}}\left( x\right) }
{\delta \Sigma(s) }
=\partial _{n}^{s}G_{\Sigma }\left(s,x\right)
\partial_{n}^{s}{\Phi_{\varphi,\Sigma}}(s) \  ,
\end{equation}%
where I have used the notation (\ref{notation}) also for
$\Phi_{\varphi,\Sigma}$.  After inserting this in (\ref{meta}),
and using (\ref{-dG=delta}) the result is
\begin{equation}
\begin{split}
\frac{\delta S[\varphi, \Sigma] }{\delta \Sigma(s) } =&
\frac{1}{2} \left( \left( \partial _{\mu }
{\Phi_{\varphi,\Sigma}(x(s))}\right) ^{2} +
m^{2}{\Phi_{\varphi,\Sigma}(x(s))}^{2}\right)
 \\
& +\int_{\Sigma } d\Sigma(t) \ \varphi(t)\
\partial_{n}^{s}{\Phi}(s) \
\partial_{n}^{s}\partial_{n}^{t}G_{\Sigma }(t,s)\ .
\end{split}
\end{equation}%

The derivative $\delta  S[\varphi, \Sigma]
/\delta \varphi\left(
z\right) $ can be easily obtained from the definition (\ref{S}).
The Hamilton-Jacobi equation (\ref{h-j}) finally reads %
\begin{equation}
\begin{split}
& \int_{\Sigma }d\Sigma(t)\ \partial _{n}^{s} {\Phi}(z)\
\varphi(t)\ \partial_{n}^{s}\partial_{n}^{t}G_{\Sigma }(t,s) +
\frac{1}{2}\left(\partial_{n}^s{%
\Phi}(s) \right) ^{2}+ \\
&\ \ \  +\frac{1}{2}\int_{\Sigma}d\Sigma(t) d\Sigma(u)\
\varphi(t)\ \varphi(u) \
\partial_{n}^{t}\partial_{n}^{s}G_{\Sigma }(s,t)\
\partial_{n}^{u}\partial_{n}^{s}G_{\Sigma }(s,u) =0\ .
\end{split}
\end{equation}%
Calculating $\partial_{n}{\Phi}(s) $ from (\ref{classica})
it is easy to show that this is an identity.

\section{Generalized Tomonaga-Schwinger equation}

It is possible to write the propagator (\ref{KK}) in
the form%
\begin{equation}
W[\varphi,\Sigma]=\sqrt{\det G_{\Sigma}}\ \exp
\left( -S[\varphi,\Sigma]] \right) =
\exp \left(
\frac{1}{2}tr\ln G_{\Sigma }-S[\varphi,\Sigma] \right) \ .
\end{equation}%
The variation with respect to the surface (dropping the arguments
from the notation) is
\begin{equation}
\frac{\delta W}{\delta \Sigma(s) }
=W \frac{\delta }{\delta
\Sigma(s) }\left(\frac{1}{2}tr\ln G_{\Sigma }-
S  \right) \  ;
\end{equation}%
so that the generalized Tomonaga-Schwinger equation (\ref{schr})
becomes%
\begin{equation}
\frac{1}{2}\frac{\delta (tr\ln G_{\Sigma }) }{\delta \Sigma (s) }-%
\frac{\delta S}{\delta \Sigma (s) }
=-\frac{1}{2}\frac{\delta ^{2}S }{\delta
\varphi(s) \delta \varphi(s)  }+\frac{1}{2}\frac{\delta S}{%
\delta \varphi(s)}\frac{\delta S}
{\delta \varphi(s)}-\frac{1%
}{2}( \nabla \phi(s)) ^{2}-\frac{1}{2}m^{2}\phi ^{2}(s)\  .
\end{equation}%
Using the Hamilton-Jacobi equation (%
\ref{h-j}) derived in the last section, this equation reduces to%
\begin{equation}
\frac{\delta (tr\ln G_{\Sigma }) }{\delta \Sigma (s) }=-%
\frac{\delta ^{2}S}{\delta
\varphi(s) \delta \varphi(s) }\  .
\end{equation}%
The left hand side of this equation can be calculated again using
Hadamard's formula (\ref{hadamard}) and (\ref%
{defG})%
\begin{equation}
\frac{\delta tr\ln G_{\Sigma }}{\delta \Sigma (s) }=%
tr\left[ \left( -\square _{x}+m^{2}\right) \frac{\delta G_{\Sigma
}\left( x,y\right) }{\delta \Sigma (s) }\right] =%
\left. \partial_{n}^{t}\partial_{n}^{u}G_{\Sigma }\left(
z_{1},z_{2}\right) \right\vert_{t=u=s}\  ,
\end{equation}%
while the right hand side can be directly obtained from (\ref{S}),
giving
\begin{equation}
\frac{1}{2}\frac{\delta ^{2}}{\delta
\varphi(s) \delta \varphi(s) }\
\int_{\Sigma} d\Sigma(s) d\Sigma(t) \ \varphi(s)\
\varphi(t) \ \partial_{n}^{s}\partial_{n}^{t}G_{\Sigma
}(s,t)=
\left. \partial_{n}^{t}\partial_{n}^{u}G_{\Sigma }\left(
z_{1},z_{2}\right) \right\vert_{t=u=s}\  ,
\end{equation}%
showing that the equation is satisfied.

\section{Conclusions}

I have derived the generalized Tomonaga--Schwinger equation
(\ref{equazione}) from its functional integral definition
(\ref{K}), in the case of a free euclidean scalar field.

An issue that I have not investigated is given by possible
restrictions on the shape of the region $\cal R$ in using Green
functions techniques; for example whether it has to be convex.

I think that this line of investigation should be relevant in
order to develop the program presented in \cite{cdort} to make
contact between background independent quantum gravity and
conventional calculations of particle scattering amplitudes.

The result derived here undercuts some a priori arguments against
the general possibility of an equation as the form
(\ref{equazione}), and renders the equation more plausible.
However, it is of course far from a general proof of the validity
of this equation.  In the lorentzian case, in particular, the
distinction between initial and final configurations on the one
hand, and boundary values of the fields on the other, might
appear more substantial.  The possibility that the variation of
the two be governed by the same equation is less evident.  The
difficulty of using functional integral techniques is that
integrals that converge in the euclidian oscillate in the
lorentzian case.  If convergence of the integrals can be
controlled, I expect that it should be possible to prove the
validity of the generalized Tomonaga--Schwinger using the same
strategy as here.

\section*{Acknowledgements}

It is my pleasure to express warm thanks to Professor Carlo
Rovelli, for having suggested the subject of this research and for
his constant generous help and encouragement, and to Professor
Massimo Testa for many discussions and invaluable comments and
suggestions, among which the idea of using Hadamard's equation,
which have been essential for the completion of this study.

\appendix

\section*{Appendix}

\subsection*{A: Hadamard formula}

I sketch here the demonstration of the Hadamard formula
(\ref{hadamard}).  Consider two surfaces $\Sigma _{1}$ and $\Sigma
_{2}$, where $\Sigma _{2}$ coincides with $\Sigma _{1}$ except
for an infinitesimal outward bulge around a point. Consider the difference $%
G_{\Sigma _{2}}\left( x,y\right) -G_{\Sigma _{1}}\left(
x,y\right) $. This difference satisfies the equation of motion
(\ref{eqmoto}). If $x\in \Sigma_{2}$:
\begin{equation}
\left. G_{\Sigma _{2}}\left( x,y\right) -G_{\Sigma _{1}}\left(
x,y\right) \right\vert _{x\in \Sigma _{2}}=-G_{\Sigma _{1}}\left(
x,y\right) \  ;
\end{equation}%
taking $y\in \Sigma $ in (\ref{classica}) shows that%
\begin{equation}
-\partial _{n}^{x}G_{\Sigma }\left( x,y\in \Sigma \right) =\left. \delta
^{\left( 3\right) }\left( x-y\right) \right\vert _{y\in \Sigma }\  ,
\label{-dG=delta}
\end{equation}%
so that%
\begin{equation}
G_{\Sigma _{2}}\left( x,y\right) -G_{\Sigma _{1}}\left( x,y\right)
=\int_{\Sigma _{2}}d\Sigma _{2}(s) \partial _{n}^{z}G_{\Sigma
_{2}}\left( z,x\right) G_{\Sigma _{1}}\left( z,y\right) \  .
\end{equation}%
The region of $\Sigma _{2}$ where $\Sigma _{2}\equiv \Sigma _{1}$
doesn't contribute to this integral because of (\ref{GxeS=0}). In
the region
where $%
\Sigma _{2}\neq \Sigma _{1}$, $G_{\Sigma _{1}}\left( z,y\right) $ is
infinitesimal, so it is possible to approximate%
\begin{equation}
G_{\Sigma _{1}}\left( z,y\right) \simeq \partial _{n}^{z}G_{\Sigma
_{2}}\left( z,y\right) \delta n\  ,
\end{equation}%
where $n$ is the normal to $\Sigma _{1}$, positive in the outgoing
direction. As a consequence we have Hadamard's formula:%
\begin{equation}
\frac{\delta G_{\Sigma }\left( x,y\right) }{\delta \Sigma (s) }%
=\partial _{n}^{z}G_{\Sigma }\left( z,x\right) \partial
_{n}^{z}G_{\Sigma }\left( z,y\right) \  .
\end{equation}

\subsection*{B: The field propagator}

Here I give the explicit expression for the propagator
$W[\varphi,\Sigma]$ in the case in which $\Sigma$ is formed by two
parallel planes $x^4\equiv t=0$ and $x^4 \equiv t=T$.  That is, I
give the functional field propagator of the free euclidean scalar
theory.  I denote $\varphi_{1}(\vec x)$ and $\varphi_{2}(\vec x)$,
the value of the field on the initial and final surface,
respectively, where $\vec x = (x^1,x^2,x^3)$ are cartesian
coordinates on the planes, and I denote the propagator as
$W[\varphi_{1},\varphi_{2},\ T]$.  The gaussian integral
($\ref{K}$) can be solved by finding the extremal value of the
exponent, that is, by solving the classical equation (\ref
{eqmoto}) with boundary conditions $\Phi(\vec x,0)=\varphi_1(\vec
x)$
and  $\Phi(\vec x,T)=\varphi_2(\vec x)$.  The solution is%
\begin{equation}
{\Phi}(\vec x,t)=\int \frac{d^{3}kd^{3}y}{( 2\pi) ^{3}}%
\ e^{-i\vec{k}( \vec{x}-\vec{y}) }\ \frac{\varphi_{2}(\vec{y}%
)\ \sinh \omega t-\varphi_{1}( \vec{y})\ \sinh \omega( t-T)
}{\sinh \omega T}\       .
\end{equation}%
Inserting this in the action and computing the gaussian integral
gives
\begin{equation}
\begin{split}
& W[\varphi_{1},\varphi_{2},T]=N\exp \left\{ -\frac{1}{2}%
\int_{-\infty }^{+\infty }d^{3}x\int_{-\infty }^{+\infty }d^{3}y\int
\frac{%
d^{3}k}{\left( 2\pi \right) ^{3}}\ \ e^{-i\vec{k}\left(
\vec{y}-\vec{x}\right) }%
\ \ \sqrt{\vec{k}^{2}+m^{2}}\hspace{3em}\right.  \\ & \ \ \ \left.
\left( -\frac{\phi _{1}\left( \vec{y}\right) \phi _{2}(\vec{x})
+\phi _{1}(\vec{x}) \phi _{2}(\vec{y}) }{\sinh\!\big(
\sqrt{\vec{k}^{2}+m^{2}}T\big) }+\coth\!
\big(\sqrt{%
\vec{k}^{2}+m^{2}}T\big) \left(\phi _{2}(\vec{y}) \phi
_{2}(\vec{x}) +\phi _{1}(\vec{y}) \phi_{1}(
\vec{x}) \right) \right)\right\}   .
\end{split}
\end{equation}%
The normalization factor $N$ can be found via the Schr\"{o}dinger
equation; the solution is%
\begin{equation}
N\propto \exp \left( \frac{V}{2}\int \frac{d^{3}k}{\left( 2\pi \right)
^{3}}%
\ln \sinh \omega T\right)
\end{equation}%
with $V$ being the spatial volume.  For completeness, I give also
the propagator in the Lorentzian case
\begin{equation}
\begin{split}
&W[\varphi_{1},\varphi_{2},T]=N\exp \left\{ \frac{i}{2}%
\int_{-\infty }^{+\infty }d^{3}x\int_{-\infty }^{+\infty }d^{3}y\int
\frac{%
d^{3}k}{\left( 2\pi \right) ^{3}}\ \ e^{-i\vec{k}\left(
\vec{y}-\vec{x}\right) }%
\ \ \sqrt{\vec{k}^{2}+m^{2}}\hspace{3em}\right.  \\ & \ \ \
 \left. \left( -\frac{\phi _{1}\left( \vec{y}\right) \phi
_{2}(\vec{x}) +\phi _{1}(\vec{x}) \phi
_{2}(\vec{y}) }{\sin\!\big( \sqrt{\vec{k}^{2}+m^{2}}T\big) }+\cot\!
\big(\sqrt{%
\vec{k}^{2}+m^{2}}T\big) \left(\phi _{2}(\vec{y}) \phi
_{2}(\vec{x}) +\phi _{1}(\vec{y}) \phi_{1}( \vec{x}) \right)
\right)\right\}   .
\end{split}
\end{equation}%

\end{document}